\documentclass{article}
\usepackage{amssymb,color,graphicx}
\usepackage{amssymb}
\usepackage{wrapfig,framed,cancel}
\usepackage[colorlinks=true, pdfstartview=FitV, linkcolor=blue, citecolor=black, urlcolor=blue]{hyperref}
\definecolor{shadecolor}{rgb}{0.93, 0.93, 0.86}
\newtheorem{theorem}{Theorem}[section]
\newtheorem{prop}{Proposition}[section]

\newtheorem{remark}{Remark}[section]
\newtheorem{problem}{Problem}[section]

\def\br{\begin{remark}}
\def\er{\end{remark}}

\def\C{{\mathbb C}}

\def\bR{{\mathbb R}}
\def\eqref#1{ (\ref{#1})}
\def\&{&\hspace{-20pt}}
\def\R{{\mathbb R}}
\def\d{{\rm d}}

\def\1{\mathbf 1}

\def\wt{\widetilde}
\def \pa{\partial}

\def\bea{\begin{eqnarray}}
\def\eea{\end{eqnarray}}
\def\Id{\mathrm{Id}}

\def\f{{\bf f}}
\def\g{{\bf g}}
\def\vf{\vec{f}}
\def\vg{\vec{g}}
\def\F{{\mathcal F}}
\def\G{{\mathcal G}}
\def\H{{\mathcal H}}
\def\0{{\bf 0}}

\def\res{\mathop{{\rm res}}}
\def\Tr{\mathop{{\rm Tr}}}

\def\T{\mathrm{T}}
\def\bC{{\bf C}}
\def\bD{{\bf D}}

\def\QED{ {\bf Q.E.D}\par \vskip 4pt}
\def\K{{\mathfrak K}}
\newcommand{\be}{\begin{eqnarray}}
\newcommand{\ee}{\end{eqnarray}}
\newcommand{\bes}{\begin{eqnarray*}}
\newcommand{\ees}{\end{eqnarray*}}
\newcommand{\ds}{\displaystyle}

\definecolor{light-blue}{rgb}{0.8,0.85,1}
\definecolor{blue}{rgb}{0,0,1}
\definecolor{red}{rgb}{1,0,0}

\def\le{\left}
\def\ri{\right}
\def\ba{\begin{eqnarray}}
\def\eeq{\end{eqnarray}}

\renewcommand{\theequation}{\arabic{section}.\arabic{equation}}
\makeatletter
\@addtoreset{equation}{section}
\makeatother

\begin{document}

\baselineskip 16pt plus 1pt minus 1pt
\begin{flushright}
\end{flushright}
\vspace{0.2cm}
\begin{center}
\begin{Large}
\textbf{Riemann--Hilbert approach to multi--time processes; the Airy and the Pearcey case.}
\end{Large}

\bigskip
M. Bertola$^{\dagger\ddagger}$\footnote{Work supported in part by the Natural
  Sciences and Engineering Research Council of Canada (NSERC)}\footnote{bertola@crm.umontreal.ca},  
M. Cafasso$^{\dagger\ddagger}$ \footnote{cafasso@crm.umontreal.ca}
\\
\bigskip
\begin{small}
$^{\dagger}$ {\em Centre de recherches math\'ematiques,
Universit\'e de Montr\'eal\\ C.~P.~6128, succ. centre ville, Montr\'eal,
Qu\'ebec, Canada H3C 3J7} \\
\smallskip
$^{\ddagger}$ {\em  Department of Mathematics and
Statistics, Concordia University\\ 1455 de Maisonneuve W., Montr\'eal, Qu\'ebec,
Canada H3G 1M8} 
\\
\end{small}
\end{center}
\bigskip
\begin{center}{\bf Abstract}\\
\end{center}
We prove that matrix Fredholm determinants related to multi--time processes can be expressed in terms of determinants of integrable kernels \`a la Its-Izergin-Korepin-Slavnov (IIKS) and hence related to suitable Riemann-Hilbert problems, thus extending the known results for the single-time case. We focus on the Airy and Pearcey processes.  As an example of applications we re-deduce a third order PDE, found by Adler and van Moerbeke, for the two--time Airy process. 
\tableofcontents

\section{Introduction and description of results}

Dyson, in \cite{Dy:BrownianMotions}, described how to implement a dynamics into random matrix models in such a way that the eigenvalues of the matrix behave like finitely many non--intersecting Brownian motions on the real line. In a suitable scaling regime we are lead to the study of certain interesting time--dependent determinantal random point processes generalizing the typical probability distributions appearing in random matrix models (see \cite{TW-Dyson}). This paper is about two of these processes, described respectively by the matrix Airy and Pearcey kernels (see  (\ref{AKern})--(\ref{gaussian}) and (\ref{PKern})--(\ref{gaussianp})). The Airy process was introduced in \cite{PrSp} and further developed in \cite{JoDetProc,Johansson3}, while the Pearcey process was introduced in \cite{TracyWidomPearcey,OkounkovReshetikhin} in the context of non-intersecting Brownian motions and plane partitions. 

One important feature of the probability distributions appearing in random matrices it is that they are related to Fredholm determinants of integrable operators \`a la Its--Izergin--Korepin--Slavnov \cite{ ItsIzerginKorepinSlavnov}. Namely,   their kernels can be written in the form
\be
	K(x,y)=\frac{\vec{f}^\T(x)\vec{g}(y)}{x-y}
\ee 
where $\vec{f},\vec{g}$ are two vectors of a given dimension such that $\vec{f}^\T(x)\vec{g}(x)=0$ (so that, in particular, the kernel is non--singular on the diagonal). This simple fact has important consequences, as it reduces the computation of the Fredholm determinant to the study of a certain Riemann--Hilbert boundary--value problem canonically related to the operator's kernel (for a concise account see \cite{ItsHarnad} and also the appendix A below). If, as it is very often the case, the determinant cannot be written in terms of elementary functions, still the Riemann--Hilbert formulation is very useful for finding some differential equations and studying asymptotic properties of the determinant.
In this paper we will prove that the determinants of the Airy and the Pearcey matrix operators (namely the multi-time gap probabilities of the respective point processes) are equal to the determinants of some explicitly given integrable kernels (for details see Theorems \ref{mainAiry} and \ref{mainPearcey})
\bea
		\mathrm{det}(\Id-\chi_{\vec I}A)&=&\mathrm{det}(\Id-K_A) \\
		\mathrm{det}(\Id-\chi_{\vec J}P)&=&\mathrm{det}(\Id-K_P)
\eea
where $A,P$ are respectively the matrix Airy and Pearcey operator, $\chi_I$ and $\chi_J$ are the indicator functions of a given collection of intervals and $K_A, K_P$ are of the form
\bea
	K_A(\lambda,\mu)=\frac{\f_A^\T(\lambda)\g_A(\mu)}{\lambda-\mu}\ ,\qquad
	K_P(\lambda,\mu)=\frac{\f_P^\T(\lambda)\g_P(\mu)}{\lambda-\mu}.
\eea
Now $\f$ and $\g$, for both cases, will be rectangular matrices of appropriate size, as in \cite{ItsHarnad}. Our approach is the same used in \cite{BertolaCafasso1} for the scalar Airy and Pearcey operators. We believe that this method, with appropriate modifications, can lead to the proof that also others matrix kernels (like the Hermite, the sine and the Bessel described in \cite{TW-Dyson}) are of integrable type. As an example of possible applications we describe how to obtain a system of isomonodromic Lax equations for the Airy and Pearcey processes. In the simplest case of the two--time Airy process on two semi--infinite intervals, we also re--derive from the Lax pair a partial differential equation originally discovered by Adler and van Moerbeke \cite{AvM-Airy-Sine} and generalized for an arbitrary number of times by Dong Wang in \cite{DongWang}.

Having a Riemann--Hilbert formulation for these Fredholm determinants can be used
to address asymptotic of the multi-time Pearcey gap probabilities and how it connects with multi-time Airy gap probabilities, along the lines \cite{BertolaCafasso1}. 

\section{The Airy kernel}
\label{AiryKernel}
The multi-time Airy process with times $\tau_1<\tau_2<\ldots<\tau_n$ is governed by the matrix Fredholm operator $A := \wt A - B$ with kernels $A(x,y), \wt A(x,y), B(x,y)$ whose $(i,j)$-entries are given by \cite{JoDetProc}\footnote{Here the contours are slightly modified with respect to the one used in the reference. Still, our contours can be deformed into the original ones leaving the kernel of the operator unchanged.}
\bea
A_{ij}(x,y) &:=&\tilde A_{ij}(x,y)-{B_{ij}(x,y)}\ , 1\leq i,j \leq n
\label{Adef}\\
\wt A_{ij}(x,y)&:=&\frac{1}{(2\pi i)^2} \int_{\gamma_{R_i}}\hspace{-10pt} \d\mu \int_{i\bR} \d \lambda \frac {{\rm e}^{\theta(x,\mu) - \theta(y,\lambda)}}{\lambda + \tau_j - \mu - \tau_i} \label{AKern}\\
\theta(x,\mu)&:=& \frac {\mu^3}3  -  x\mu.\label{thetaAi}\\
B_{ij}(x,y) &:=& \chi_{\tau_i<\tau_j} \frac{1}{\sqrt{4\pi(\tau_j-\tau_i)}}{\rm e}^{\frac{(\tau_j-\tau_i)^3}{12}-\frac{(x-y)^2}{4(\tau_j-\tau_i)}-\frac{(\tau_j-\tau_i)(x+y)}2}\label{gaussian}
\eea


Note that the matrix $B_{ij}(x,y)$ is strictly upper triangular.
Here $\gamma_{R_i}:=\gamma_R-\tau_i$ and $\gamma_R$ consists of two oriented rays, one from $\infty {\rm e}^{\frac{\pi i}3}$ to $C$ and the other from $C$ to $\infty {\rm e}^{-\frac{\pi i}3}$. $C$ is a real number chosen so that $C>\max_j\{\tau_j\}$. Consider the collection of multi--intervals $\lbrace I_1,\ldots, I_n \rbrace$ where 
\bea
I_j:=\le\{\begin{array}{cc}
[a_j^{(1)},a_j^{(2)}]\cup[a_j^{(3)},a_j^{(4)}]\cup\ldots\cup [a_j^{(k_j-1)},a_j^{(k_j)}] & \hbox{ if }k_j \hbox{ is even}\\[1pt]
[a_j^{(1)},a_j^{(2)}]\cup[a_j^{(3)},a_j^{(4)}]\cup\ldots\cup [a_j^{(k_j)},\infty) & \hbox{  if  } k_j \hbox{  is odd}.
\end{array}
\ri.
\eea
 Moreover denote with $\chi_{\vec I}$ its characteristic function. In the following, given any function $h$, we denote with $h(\vec a_j)$ the column vector whose $i^{th}$ component is $h(a_j^{(i)})$. Also we denote $\epsilon_k:=\mathrm{diag}(1,-1,1,-1,\ldots,(-1)^{k+1})$. The main result of the section is the following:

\begin{theorem}\label{mainAiry} The following identity between Fredholm determinants holds
	\be
		\mathrm{det}(\Id-\chi_{\vec I}A)=\mathrm{det}(\Id-K_A)
	\ee
	where $K_A$ is the integrable Fredholm operator acting on $H:=(\bigoplus_{i=1}^nL^2(i\bR+\tau_i,\C^n))\oplus L^2\le(\gamma_R,\C^n\ri)$ with kernel
	\be
		&&K_A(\lambda,\mu)=\frac{\f_A^\T(\lambda)\g_A(\mu)}{\lambda-\mu}\label{defK_A1}\\
		&&\f_A:=\frac{1}{2\pi i}(\vec f_{A,1},\ldots,\vec f_{A,n}),\quad \g:=(\vec g_{A,1},\ldots, \vec g_{A,n})\label{defK_A2}\\
		&&\vf_{A,i}(\lambda):=\le[\begin{array}{c}
								{\rm e}^{\frac{1}2\theta(0,\lambda_j)}\chi_{\gamma_R}(\lambda)\\
								\\
								\0_{k_1}\\
								\\
								\vdots\\
								\\
								\0_{k_{i-1}}\\
								\\
								{\rm e}^{\vec a_i\lambda_i}\chi_{i\bR+\tau_i}(\lambda)\\
								\\
								\0_{k_{i+1}}\\
								\\
								\vdots\\
								\\
								\0_{k_n}
							\end{array}\ri],\quad
				\vg_{A,j}(\mu):=\le[\begin{array}{c}
								{\rm e}^{-\theta(0,\mu_j)}\chi_{i\bR+\tau_j}(\mu)\\
								\\
								\epsilon_{k_1}\;{\rm e}^{\theta(\vec a_1,\lambda_1)-\theta(0,\lambda_j)}\chi_{i\bR+\tau_j}(\mu)\\
								\\
								\vdots\\
								\\
								\epsilon_{k_{j-1}}\;{\rm e}^{\theta(\vec a_{j-1},\lambda_{j-1})-\theta(0,\lambda_j)}\chi_{i\bR+\tau_j}(\mu)\\
								\\
								\epsilon_{k_j}\;{\rm e}^{\frac{1}2\theta(0,\mu_j)-\vec a_j\mu_j}\chi_{\gamma_R}(\mu)\\
								\\
								\0_{k_{j+1}}\\
								\\
								\vdots\\
								\\
								\0_{k_n}
							\end{array}\ri]
	\label{defK_A3}
\ee
Here and below the variables $\lambda,\mu,\xi$ with index, like for instance $\lambda_i$, denote $\lambda_i:=\lambda-\tau_i$.
\end{theorem}
\begin{remark}
\label{caveatAiry}
The naming of Fredholm determinant in Thm. \ref{mainAiry} is slightly abusive: strictly speaking here  `` $\det$'' is defined through the Fredholm expansion (\ref{FredhExp}). In more precise terms
the operator $\chi_{\vec I} \tilde A$ is of trace-class,  whereas $\chi_{\vec I} B$ is Hilbert--Schmidt (this will appear clear in the proof below after (\ref{matrixK_A})) and its kernel is patently diagonal--free. Therefore 
\be
``\det"(\Id - A) :=
``\det"(\Id - \wt A + B) ={\rm e}^{-\Tr \wt A}
 \mathrm{det}_2(\Id - \wt A + B)
 \ee
 which is the functional-analytically sound definition. Here $\det_2$ denotes the Carleman regularized definition of determinant, for which we refer to Appendix \ref{appregdet}.
\end{remark}
\noindent\emph{Proof:} We work on the entry $(i,j)$ of the kernel $A(x,y)$ (\ref{Adef}) and we observe that we can write (using Cauchy's theorem and after an appropriate shift of the variables of integration)
\be
&& A_{ij}(x,y)\chi_{I_i}(x)=\nonumber\\
&& \int_{i\bR+\tau_i}\frac{d\xi}{2\pi i}\sum_{\ell=1}^{k_i}(-)^{\ell+1}e^{\xi_i(a_i^{(\ell)}-x)}\times\nonumber\\
&&\left[\int_{i\bR+\tau_j}\frac{d\lambda}{2\pi i}\int_{\gamma_R}\frac{d\mu}{2\pi i}\frac{{\rm e}^{\theta(a_i^{(\ell)},\mu_i)-\theta(0,\lambda_j)+y\lambda_j}}{(\xi-\mu)(\mu-\lambda)}+\chi_{\tau_i<\tau_j}\int_{i\bR+\tau_j}\frac{d\mu}{2\pi i}\frac{{\rm e}^{\theta(a_i^{(\ell)},\mu_i)-\theta(0,\mu_j)+y\mu_j}}{\xi-\mu}\right]\nonumber
\ee
We deduce that $\chi_{\vec I}A=\mathcal T^{-1}\K_A\mathcal T$
where $\mathcal T = {\rm diag} (\mathcal T_1,\dots , \mathcal T_n)$ is the following diagonal Fourier transform, $\mu_j:= \mu-\tau_j$
\be
& \begin{array}{rcl}
	\mathcal T_j: L^2(i\bR+\tau_j)&\longrightarrow & L^2(\bR) \\
f(\mu) & \mapsto & \ds \frac{1}{\sqrt{2i\pi}}\int_{i\bR+\tau_j}{\rm e}^{\mu_j x} f(\mu)\d\mu
\end{array}
\begin{array}{rcl}
	\mathcal T_j^{-1}: L^2(\bR)&\longrightarrow & L^2(i\bR+\tau_j) \\
h(x) & \mapsto & \ds \frac{1}{\sqrt{2i\pi}}\int_{\bR}{\rm e}^{-\mu_j x} h(x)\d x.
\end{array}
\nonumber\ee
and $\K_A$ has kernel explicitly given by entries
\be
	(\K_{A})_{ij}(\xi,\lambda)=\sum_{\ell=1}^{k_i}(-1)^{\ell+1}\left[\int_{\gamma_R}\frac{d\mu}{2\pi i}\frac{{\rm e}^{\theta(a_i^{(\ell)},\mu_i)-\theta(0,\lambda_j)+a_i^{(\ell)}\xi_i}}{(\xi-\mu)(\mu-\lambda)}+\chi_{\tau_i<\tau_j}\frac{{\rm e}^{\theta(a_i^{(\ell)},\lambda_i)-\theta(0,\lambda_j)+a_i^{(\ell)}\xi_i}}{\xi-\lambda}\right].
\label{K_A}\ee
Here $\K_A$ is an operator\footnote{It is worth noticing that $\K_A$, as written in (\ref{K_A}), is already an integrable kernel \`a la IIKS. Still, in order to have a more ``well behaved'' Riemann--Hilbert problem, we go on manipulating the kernel a little bit more.} acting on $\bigoplus_{j=1}^n L^2(i\bR+\tau_j,\C^n)$. Using (\ref{K_A}) we can write\\
$\K_A=\G_A\circ\F_A+\H_A$ where $\G_A,\F_A$ and $\H_A$ are the matrix Fredholm operators
\be 
	&&\F_A: \bigoplus_{i=1}^n L^2(i\bR+\tau_i,\C^n)\longrightarrow L^2(\gamma_R,\C^n)\quad\quad\quad\quad \G_A: L^2(\gamma_R,\C^n)\longrightarrow \bigoplus_{i=1}^n L^2(i\bR+\tau_i,\C^n)
\nonumber\\
&&\H_A: \bigoplus_{i=1}^n L^2(i\bR+\tau_i,\C^n)\longrightarrow \bigoplus_{i=1}^n L^2(i\bR+\tau_i,\C^n) \nonumber \ee
with kernels having entries given respectively by
\be
&&(\F_A)_{ij}(\mu,\lambda):=\frac{{\rm e}^{\frac{1}2\theta(0,\mu_i)-\theta(0,\lambda_j)}}{\mu-\lambda}\quad\quad\quad (\G_A)_{ij}(\xi,\mu)=\delta_{ij}\sum_{\ell=1}^{k_i}(-1)^{\ell+1}\frac{{\rm e}^{\frac{1}2\theta(0,\mu_i)-a_i^{(\ell)}(\mu_i-\xi_i)}}{\xi-\mu}\nonumber\\
\nonumber\\
&& (\H_A)_{ij}(\xi,\lambda):=\chi_{\tau_i<\tau_j}\sum_{\ell=1}^{k_i}(-1)^{\ell+1}\frac{{\rm e}^{\theta(a_i^{(\ell)},\lambda_i)-\theta(0,\lambda_j)+a_i^{(\ell)}\xi_i}}{\xi-\lambda}.
\ee
Now consider the Hilbert space $H:=(\bigoplus_{i=1}^nL^2(i\bR+\tau_i,\C^n))\oplus L^2\le(\gamma_R,\C^n\ri)$ and $K_A:H\longrightarrow H$ the matrix Fredholm operator written in the matrix form (naturally--related to the splitting of $H$ into its two main addenda)
\be
K_A = \le[\begin{array}{c|c}
0 & \F_A \\
\hline
\G_A & \H_A
\end{array}\ri].
\label{matrixK_A}
\ee
It is easily verified by the convergence of the $L^2$ norm of their respective kernels that all the operators $\F_A,\G_A,\H_A$ are Hilbert--Schmidt and thus  $\chi_{\vec I} \wt A  = \mathcal T^{-1} \G_A\circ \F_A \mathcal T$ is of trace-class being unitarily equivalent to the product of two HS operators. On the other hand and by the same argument $\chi_{\vec I} B =  -\mathcal T^{-1} \H_A \mathcal T$ is provably only Hilbert--Schmidt. 

To avoid confusion we shall temporarily denote the determinant {\em defined} by  the Fredholm expansion (\ref{FredhExp})   by $``\det"$.
Then  $``\det"(\Id - K_A) = \det_2(\Id - K_A)$ since $\F_A,\G_A$ and $\H_A$ all have diagonal-free kernels. 

Let's also define an other operator $K'_A$ written as
\be
K'_A := \le[\begin{array}{c|c}
0 & -\F_A \\
\hline
0 & 0
\end{array}\ri].
\ee
This operator $K'_A$ is only Hilbert--Schmidt (not trace-class) and hence some care needs to be paid when inserting it into a  (regularized) Fredholm determinant. To this end we start observing that  its Carleman determinant $\det_2$ (see  App. \ref{appregdet}) is well defined and $\det_2(\Id - K'_A) \equiv 1$. Then we   have the following equalities (the operator $\tilde A$ is defined in (\ref{Adef})):
\be
	``\det"(\Id-K_A)=\mathrm{det}_2(\Id-K_A)=\mathrm{det}_2(\Id-K_A)\mathrm{det}_2(\Id-K'_A)
	=\nonumber\\
	 =\mathrm{det}_2(\Id-\K_A) {\rm e}^{-\mathrm{Tr}(\G_A\circ\F_A)}
	=\mathrm{det}_2(\Id-\chi_{\vec I}A){\rm e}^{-\mathrm{Tr}\tilde A}=``\det"(\Id-\chi_{\vec I}A).
\ee
The last equality in the first line is just an application of (\ref{det_2formula}); then we go from the first line to the second using the unitary Fourier transform $\mathcal T$ and the last equality is due to the fact that the kernel of the operator $A-\tilde A$ is diagonal-free (see Remark \ref{caveatAiry}). It just remains to prove that $K_A$ has kernel defined by the equations (\ref{defK_A1})--(\ref{defK_A3}). In view of (\ref{matrixK_A}) and definitions of $\F_A,\G_A,\H_A$ this is just a matter of straightforward computations. \QED

\subsection{Riemann--Hilbert problem for the Airy process}

Here we show how, using Theorem \ref{mainAiry}, we can relate the computation of the Fredholm determinant of the matrix Airy operator to the theory of isomonodromic equations. We start defining the following RH problem that, according to the IIKS theory, is naturally related to the Fredholm determinant $\det(\Id-K_A)$.

\begin{problem}\label{AiRH}

Let $\Gamma(\lambda)$ be the sectionally analytic function (if existing) that solves the RHP
\be
&&\Gamma_+(\lambda) = \Gamma_-(\lambda) (\1-2\pi iG_A(\lambda))\ ,\quad\quad \lambda\in\gamma:= \gamma_R\cup\bigcup_{i=1}^n(i\bR+\tau_i)\\
&&\Gamma(\lambda) = 
\le(\1 + \sum_{j=1}^{\infty} \frac {\Gamma_j}{\lambda^j}\ri),\quad\quad \lambda\longrightarrow\infty\\
&&G_A(\lambda):= 
				 \le[\begin{array}{ccccc}
					0 & (\epsilon_{k_1}\;{\rm e}^{\theta(\vec a_{1},\lambda_1)})^\T\chi_{\gamma_R} & (\epsilon_{k_2}\;{\rm e}^{\theta(\vec a_{2},\lambda_2)})^\T\chi_{\gamma_R} & \ldots & (\epsilon_{k_n}\;{\rm e}^{\theta(\vec a_{n},\lambda_n)})^\T\chi_{\gamma_R}\\
					&&&&\\
					{\rm e}^{-\theta(\vec a_{1},\lambda_1)}\chi_{i\bR+\tau_1} & 0 & \ldots & \ldots & 0\\
					&&&&\\
					{\rm e}^{-\theta(\vec a_{2},\lambda_2)}\chi_{i\bR+\tau_2} & \bC_{2,1}(\lambda) & \ddots & \vdots & \vdots\\
					&&&&\\
					\vdots & \vdots & \ddots & \ddots & \vdots\\
					&&&&\\
					{\rm e}^{-\theta(\vec a_{n},\lambda_n)}\chi_{i\bR+\tau_n} & \bC_{n,1}(\lambda) & \ldots & \bC_{n,n-1} & 0
					\end{array}\ri]\nonumber\\
					&&\nonumber\\
		&&\bC_{i,j}(\lambda)=\chi_{i\bR+\tau_i}(\lambda)\left[(-1)^{t+1}{\rm e}^{-\theta(a_i^{(s)},\lambda_i)_+\theta(a_j^{(t)},\lambda_j))}\right]_{s=1\ldots k_i, t=1\ldots k_j}\hspace{-15pt}\in\mathrm{Mat}(k_i\times k_j)
\ee
\end{problem}
\begin{remark}
The jump on the contours $i\bR+\tau_j$  is oscillatory on the first column due to the term ${\rm e}^{-\theta(\vec a_j,\lambda_j)}$ while on the other columns is exponentially decaying (this is due to the fact that the leading term of the exponential in $\bC_{i,j}$ is $3(\tau_i-\tau_j)\lambda^2$ with $i<j$). In order to have exponential decay in all columns we can now deform all the contours $i\bR+\tau_j$ to $\gamma_L+\tau_j$ with $\gamma_L:=-\gamma_R$. We could also have collected all these new jumps on the single contours $\gamma_L$  by multiplication of the jumps. The resulting jump, however, would be slightly more complicated.

\end{remark}

It is straightforward that $G_A(\lambda)=\f_A(\lambda)\g_A^\T(\lambda)$ and that $\g_A^T(\lambda) \f_A(\lambda) \equiv \0_n$. Hence, using Theorem \ref{thdetMalgrange}, we arrive to the following

\begin{theorem}\label{Fredholmtau}
The Fredholm determinant $\det(\Id-\chi_{\vec I}A)$ is equal to the isomonodromic tau function $\tau_{JMU}$ in (\ref{taufunction}) related to the RH problem (\ref{AiRH}). In particular $\forall\, i=1,\ldots, n$ and $\forall \,\ell=1,\ldots, k_i$ we have
\be
	&&\pa  \ln \det(\Id-\chi_{\vec I}A) = \int_{\gamma}\mathrm{Tr}\Big(\Gamma_-^{-1}(\lambda )\Gamma'_-(\lambda)\Xi_{\partial }(\lambda )\Big)\frac{\d\lambda}{2\pi i}\nonumber\\
	&&\Xi_{\partial}(\lambda ):= \le(\partial (\1- 2 \pi i\, G_A\ri)(\1 - 2\pi i\, G_A)^{-1}
	 = - 2 \pi i\,\pa  G_A(\1 + 2\pi i\, G_A) \nonumber
\ee
where we have denoted by $\pa$ any of the derivatives $\pa_{a_i^{(\ell)}}, \pa_{\tau_i}$.
\end{theorem}

Theorem \ref{Fredholmtau} implies also some more explicit differential identities by using the Miwa-Jimbo-Ueno residue formula; note first that the jump matrices $G_A$ can be written as 
\bea
G_A(\lambda) = {\rm e}^{T_A(\lambda)} G_A^{(0)} {\rm e}^{-T_A(\lambda)}
\eea
where $G_A^{(0)}$ is a {\em constant} matrix (consisting of only $\pm 1$ and $0$) and 

\bea
&&T_A(\lambda) = {\rm diag}\le(T_A^{(0)},\vec T_A^{(1)},\dots, \vec T_A^{(n)}\ri),\qquad T_A^{(0)} := \frac{1}{1+\sum_{i=1}^n k_i} \sum_{i=1}^{n}\sum_{\ell=1}^{k_i} \theta(a_i^{(\ell)},\lambda_i)\label{TA}\\
&& \vec T_A^{(i)} := T_A^{(0)}\1-{\rm diag}\le(\theta(a_i^{(1)},\lambda_i),\ldots,\theta(a_i^{(k_i)},\lambda_i)\ri)\ ,\qquad  \Tr \,\,T_A(\lambda)\equiv 0.
\eea
Therefore the matrix $\ds \Psi_A(\lambda):= \Gamma_A(\lambda)\, {\rm e}^{T_A(\lambda)} $
 solves a RHP with constant jumps and hence is (sectionally) a solution to a polynomial ODE
 \bea
 \pa_\lambda  \Psi_A(\lambda) = L(\lambda) \Psi_A(\lambda)\ ,\ \ \ \deg L(\lambda) = 2. \label{Lax}
 \eea
 
   It was shown in \cite{BertolaIsoTau} and it is not hard to see using Cauchy's residue theorem (see also \cite{BertolaCafasso2}) 
   that 
   \be
 \int_{\gamma}\mathrm{Tr}\Big(\Gamma_-^{-1}(\lambda)\Gamma'_-(\lambda)\Xi_{\partial }(\lambda)\Big)\frac{\d\lambda}{2\pi i} = -  \res_{\lambda=\infty} {\rm Tr} \le(\Gamma^{-1}(\lambda) \Gamma'(\lambda) \pa T_A \ri)\d \lambda \label{219}
\ee
where the (formal) residue in (\ref{219}) simply stands for minus the coefficient of the power $\lambda^{-1}$ in the asymptotic expansion of the argument.
Direct application of (\ref{219}) using the expression of $T_A$ (\ref{TA}) and Theorem \ref{Fredholmtau} yields the following proposition.
\begin{prop}
\label{Freddef}
The Fredholm determinant $\det(\Id-\chi_{\vec I}A)$ satisfies
\bea
\pa_{a_i^{(\ell)}} \ln \det(\Id-\chi_{\vec I}A) =-(\Gamma_1)_{1+(\sum_{j<i}k_j)+\ell,1+(\sum_{j<i}k_j)+\ell}\\
\pa_{\tau_i} \ln \det(\Id-\chi_{\vec I}A)=\sum_{\ell=1}^{k_i}(2\tau_i\Gamma_1+\Gamma_1^2-2\Gamma_2)_{1+(\sum_{j<i}k_j)+\ell,1+(\sum_{j<i}k_j)+\ell}
\eea
where $\Gamma_1, \Gamma_2$ and $\Gamma(\lambda)$ are the coefficients of the asymptotic expansion of the solution of the RHP \ref{AiRH}.
\end{prop}

\begin{remark}[On the existence of the solution to Prob. \ref{AiRH}]\label{exAi}
According to Thm.\ref{mainAiry} and following the general IIKS theory of integrable operators, it follows that the solution to Problem \ref{AiRH} exists if and only if the Fredholm determinant $\det(\Id-K_A)$ does not vanish. The latter determinant is equal to $\det(\Id-\chi_{\vec I}A)$ by Theorem \ref{Fredholmtau}, and this follows from the probabilistic interpretation of the same. Of course this should be proved also independently, for example by showing that the operator $A$  has operator norm bounded by one.
\end{remark}

\subsection{Lax formulation of a PDE for the Airy process}
Using the representation (\ref{Lax}) above we can give an independent verification for the nonlinear PDE satisfied by $G(\tau,a,b)=\log\det(\Id -\chi_{\lbrace(a,\infty),(b,\infty)\rbrace} A)$ as found by Adler--van Moerbeke in \cite{AvM-Airy-Sine} and generalized by D. Wang \cite{DongWang} to the case of an arbitrary number of times (here $\tau=\tau_2-\tau_1$):
\be
\le(\frac {\tau^2}2 \pa_W - W\pa_E \ri) \le(\pa^2_{E} - \pa^2_W \ri)G + 2\tau \pa^3_{\tau E W}G = \{ \pa^2_{EW}G, \pa^2_{E} G\}_E
\label{Avm}\ee
where $E = \frac {a+b}2$, $W = \frac {a-b}2$ and $\{f, g\}_E:= \pa_E f \, g - f \pa_E g$.
We briefly describe how to use our setup to verify the equation (\ref{Avm}); the details of the computations involve a significant amount of completely straightforward algebra and will not be reported here.
We are in the case $n=2$ and $k_1=k_2=1$ so that the RHP is $3\times 3$; since the process is stationary we can assume $\tau_1=0$ and $\tau_2=\tau>0$.
The matrix $\Psi_A(\lambda)= \Gamma_A(\lambda) {\rm e}^{T_A(\lambda)}$ solves a RHP with constant jumps, where
\be
T_A(\lambda) = {\rm diag} \le(\frac {\theta(a,\lambda) + \theta(b,\lambda-\tau)}3 , \frac {\theta(b,\lambda-\tau) - 2 \theta(a, \lambda) }3 , \frac {\theta(a,\lambda)  -  2\theta(b,\lambda-\tau) }3  \ri)
\ee
Denoting by $\Gamma_j$ the coefficient matrices in the expansion of the solution as in Problem \ref{AiRH}, one can write a set of compatible PDEs
\bea
&\pa_\lambda \Psi_A(\lambda) = L(\lambda) \Psi_A(\lambda)&\label{spectral}\\
&  \pa_W \Psi_A(\lambda)  = \mathcal W (\lambda) \Psi_A(\lambda)\ ,\qquad 
  \pa_E \Psi_A(\lambda) = \mathcal E (\lambda) \Psi_A(\lambda) ,\qquad \pa_\tau \Psi_A(\lambda) = \mathcal T (\lambda) \Psi_A(\lambda)&\label{pdes}
\eea
\begin{wrapfigure}{r}{0.3\textwidth}
\resizebox{0.3\textwidth}{!}{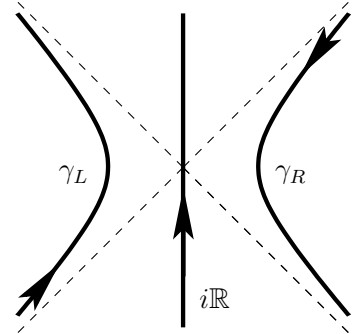}
\caption{The contours for the Pearcey kernel and Riemann--Hilbert problem}
\label{PearceyContours}
\end{wrapfigure}
with all the matrices polynomials in $\lambda$ of degrees $2,1,1,2$ respectively. From the equation (\ref{spectral}) one can express all the coefficients $\Gamma_j$ in terms of the coefficients of $\Gamma_1$ and the entries $(1,2),(2,1), (3,1),(1,3)$ of $\Gamma_2$ (we shall refer to these coefficients as the "generators"). On the other hand, equating the expansions of equations (\ref{pdes}) allows to express all derivatives in terms of these generators uniquely.
From the specialization of Prop. \ref{Freddef} we obtain 
\bea
\pa_E G = -\Tr(\Gamma_1 {\rm diag}(0,1,1))\ ,\quad\nonumber
\pa _W G = -\Tr (\Gamma_1{ \rm diag}(0,1,-1))\nonumber\\
\pa_\tau G = \Tr\le( (2\tau\Gamma_1+\Gamma_1^2-2\Gamma_2){\rm diag}(0,0,1)\ri)\nonumber
\eea
Then all the higher partial derivatives can be straightforwardly expressed in terms of polynomials in the generators.  By simply plugging  these lengthy expressions into the equation (\ref{Avm}) yields an identity. 
\begin{remark}
Although this method gives a straightforward verification of the equation, it is not {\em constructive} in that it does not provide an effective method of {\em finding} an equation. In general we would expect as many nonlinear PDEs as the number of independent variables.
\end{remark}

\section{The Pearcey kernel}
\label{PearceyKernel}
In this section, slightly modifying the approach used for the Airy kernel, we show that also the determinant of the matrix (multi-time) Pearcey operator can be expressed as a determinant of a related integrable operator. The matrix Pearcey kernel (see for example \cite{TracyWidomPearcey}) has $(i,j)$--entry given by
\bea
P_{ij}(x,y) &:=& \tilde P_{ij}(x,y)- Q(x,y,\tau_j-\tau_i)\\
\tilde P_{ij}(x,y)&:=& \frac{1}{(2\pi i)^2} \int_{\gamma_L\cup\gamma_R}\hspace{-10pt} \d\mu \int_{i\bR} \d \lambda \frac {{\rm e}^{\Theta_i(x,\mu) - \Theta_j(y,\lambda)}}{\lambda - \mu}\label{PKern}\\
\Theta_i(x,\mu)&:=& \frac {\mu^4}4 - \frac {\tau_i}2 \mu^2  -  x\mu.\label{ThetaP}\\
Q_{ij}(x,y)&:=&\chi_{\tau_i<\tau_j}  \frac{1}{\sqrt{2\pi(\tau_j-\tau_i)}}{\rm e}^{-\frac{(x-y)^2}{2(\tau_j-\tau_i)}}=\chi_{\tau_i<\tau_j}  \int_{i\bR}{\rm e}^{(\tau_j-\tau_i)\frac{\lambda^2}2+(y-x)\lambda}\frac{d\lambda}{2\pi i}\label{gaussianp}
\eea

where the contours $\gamma_L,\gamma_R$ are indicated in Fig. \ref{PearceyContours}.
Here we consider the Fredholm determinant $\det(\Id-\chi_{\vec{J}}P)$\footnote{Note that we have $\det(\Id-\chi_{\vec{J}}P)=\det(\Id-\chi_{\vec{J}}P\chi_{\vec{J}})$} where $\chi_{\vec{J}}$ is the characteristic function of the collection of multi-intervals $\lbrace J_1,\ldots, J_n \rbrace$ with $J_j:=[a_j^{(1)},a_j^{(2)}]\cup[a_j^{(3)},a_j^{(4)}]\cup\ldots\cup [a_j^{(2k_j-1)},a_j^{(2k_j)}]$ (note that for the Pearcey case we do not have semi--infinite intervals). The analog of Theorem \ref{mainAiry} is the following.
\begin{theorem}\label{mainPearcey}
 The following identity between Fredholm determinants holds
	\be
		\mathrm{det}(\Id-\chi_{\vec J}P)=\mathrm{det}(\Id-K_P)
	\ee
	where $K_P$ is the integrable Fredholm operator acting on $V:=L^2(i\bR\cup\gamma_R\cup\gamma_L)$ with kernel
	\be
		&\& K_P(\lambda,\mu)=\frac{\f_P^\T(\lambda)\g_P(\mu)}{\lambda-\mu}\label{defK_P1}\\
		&\& \f_P:=\frac{1}{2\pi i}(\vec f_{P,1},\ldots,\vec f_{P,n}),\quad \g:=(\vec g_{P,1},\ldots, \vec g_{P,n})\label{defK_P2}\\
		&\&(\vf_P)_i(\lambda):=\le[\begin{array}{c}
								{\rm e}^{\frac{1}2\Theta_i(0,\lambda)}\chi_{\gamma_L\cup\gamma_R}(\lambda)\\
								\\
								\0_{2k_1}\\
								\\
								\vdots\\
								\\
								\0_{2k_{i-1}}\\
								\\
								{\rm e}^{\vec a_i\lambda}\chi_{i\bR}(\lambda)\\
								\\
								\0_{2k_{i+1}}\\
								\\
								\vdots\\
								\\
								\0_{2k_n}
							\end{array}\ri]
				(\vg_P)_j(\mu):=\le[\begin{array}{c}
								{\rm e}^{-\Theta_j(0,\mu)}\chi_{i\bR}(\mu)\\
								\\
								\epsilon_{2k_1}{\rm e}^{-\vec a_1\mu+(\tau_j-\tau_1)\frac{\mu^2}2}\chi_{i\bR}(\mu)\\
								\\
								\vdots\\
								\\
								\epsilon_{2k_{j-1}}{\rm e}^{-\vec a_{j-1}\mu+(\tau_j-\tau_{j-1})\frac{\mu^2}2}\chi_{i\bR}(\mu)\\
								\\
								\epsilon_{2k_j}{\rm e}^{\frac{1}2\Theta_j(0,\mu)-\vec a_j\mu}\chi_{\gamma_L\cup\gamma_R}(\mu)\\
								\\
								\0_{2k_{j+1}}\\
								\\
								\vdots\\
								\\
								\0_{2k_n}
							\end{array}\ri]
	\label{defK_P3}\ee
\end{theorem}
\begin{remark}
The same 
{\em caveats} about the correct definition of $\det$ as in Remark \ref{caveatAiry} apply here: as a matter of fact, a small modification to the kernel would turn it into a fully trace-class operator (without changing the expansion of its ``Fredholm'' determinant), but for reasons of space and uniformity with the previous treatment we use the same approach.
\end{remark}

\noindent\emph{Proof:}

We work on the $(i,j)$--entry of the kernel and we observe that we can write
\be
 &\& P_{ij}(x,y)\chi_{J_i}(x)= \int_{i\bR}\frac{d\xi}{2\pi i}\sum_{\ell=1}^{2k_i}(-)^{\ell+1}e^{\xi(a_i^{(\ell)}-x)} \times\nonumber \\
&\& \times \left[\int_{i\bR}\frac{d\lambda}{2\pi i}\int_{\gamma_L\cup\gamma_R}\frac{d\mu}{2\pi i}\frac{{\rm e}^{\Theta_i(a_i^{(\ell)},\mu)-\Theta_j(0,\lambda)+y\lambda}}{(\xi-\mu)(\mu-\lambda)}+\chi_{\tau_i<\tau_j}\int_{i\bR}\frac{d\mu}{2\pi i}\frac{{\rm e}^{(\tau_j-\tau_i)\frac{\mu^2}2-a_i^{(\ell)}\mu+y\mu}}{\xi-\mu}\right]
\ee
Then 
$$\chi_{\vec J}P=\mathcal T^{-1}\K_P\mathcal T$$
where $\mathcal T$ is the Fourier transform (acting diagonally)
\be
& \begin{array}{rcl}
	\mathcal T: L^2(i\bR)&\longrightarrow & L^2(\bR) \\
f(\mu) & \mapsto & \ds \frac{1}{\sqrt{2i\pi}}\int_{i\bR}{\rm e}^{\mu x} f(\mu)\d\mu
\end{array}\quad
\begin{array}{rcl}
	\mathcal T^{-1}: L^2(\bR)&\longrightarrow & L^2(i\bR) \\
h(x) & \mapsto & \ds \frac{1}{\sqrt{2i\pi}}\int_{\bR}{\rm e}^{-\mu x} h(x)\d x.
\end{array}
\ee
and $\K_P$ has kernel explicitly given by entries
\be
	(\K_P)_{ij}(\xi,\lambda)=\sum_{\ell=1}^{2k_i}(-1)^{\ell+1}\left[\int_{\gamma_L\cup\gamma_R}\frac{d\mu}{2\pi i}\frac{{\rm e}^{\Theta_i(a_i^{(\ell)},\mu)-\Theta_j(0,\lambda)+a_i^{(\ell)}\xi}}{(\xi-\mu)(\mu-\lambda)}+\chi_{\tau_i<\tau_j}\frac{{\rm e}^{(\tau_j-\tau_i)\frac{\lambda^2}2-a_i^{(\ell)}(\lambda-\xi)}}{\xi-\lambda}\right]
\label{KP}\ee
$\K_P$ is an operator\footnote{It is worth noticing that $\K_P$ as written in (\ref{KP}) is already an integrable kernel \`a la IIKS. In particular and most importantly, there is no singularity on the diagonal $\xi=\lambda$ for the second term  thanks to the alternating sum.} acting on $L^2(i\bR,\C^n)$ and such that $\K_P=\G_P\circ\F_P+\H_P$ with
\be
	&&\F_P: L^2(i\bR,\C^n)\longrightarrow L^2\le((\gamma_L\cup\gamma_R,\C^n)\ri), \quad \G_P:L^2\le((\gamma_L\cup\gamma_R),\C^n\ri)\longrightarrow L^2(i\bR,\C^n)\nonumber\\
	&&\H_P:L^2(i\bR,\C^n)\longrightarrow L^2(i\bR,\C^n)\nonumber
\ee
and $(i,j)$--entries given respectively by
\be
	&&(\F_P)_{ij}(\mu,\lambda):=\frac{{\rm e}^{\frac{1}2\Theta_i(0,\mu)-\Theta_j(0,\lambda)}}{\mu-\lambda},\quad (\G_P)_{ij}(\xi,\mu)=\delta_{ij}\sum_{\ell=1}^{2k_i}(-1)^{\ell+1}\frac{{\rm e}^{\frac{1}2\Theta_i(0,\mu)-a_i^{(\ell)}(\mu-\xi)}}{\xi-\mu}\nonumber\\
	&&(\H_P)_{ij}(\xi,\lambda):=\chi_{\tau_i<\tau_j}\sum_{\ell=1}^{2k_i}(-1)^{\ell}\frac{{\rm e}^{a_i^{(\ell)}(\xi-\lambda)+(\tau_j-\tau_i)\frac{\lambda^2}2}}{\xi-\lambda}
\ee

Now consider the Hilbert space $V:=L^2(i\bR,\C^n)\oplus L^2\le((\gamma_L\cup\gamma_R),\C^n\ri)$ and $K_P: V \longrightarrow V$ the matrix Fredholm operator written in the matrix form (naturally--related to the splitting of $V$)
\be
K_P = \le[\begin{array}{c|c}
0 & \F_P \\
\hline
\G_P & \H_P
\end{array}\ri].
\label{matrixK_P}\ee
Here again, as in the case of the Airy operator, all the operators $\F_P,\G_P,\H_P$ are Hilbert--Schmidt so that $\mathrm{det}_2(\Id-K_P)$ is well defined. The very same argument used for the Airy operator leads us to conclude that $\det(\Id-\chi_{\vec J}P)=\det(\Id-K_P)$ (we recall once more that here ``$\det$'' stands for the Fredholm expansion (\ref{FredhExp})). Formulas (\ref{defK_P1})--(\ref{defK_P3}) are again a matter of straightforward computations using definitions of $\F_P,\G_P,\H_P$ and (\ref{matrixK_P}). \QED

\subsection{Pearcey process and isomonodromic equations}

In this section we relate the computation of the determinant of the matrix Pearcey operator to the theory of isomonodromic tau functions, thus potentially leading to a Lax representation. We start defining our RH problem
\begin{problem}\label{PRH}
Let $\Gamma(\lambda)$ be the sectionally analytic function (if existing) that solves the RHP
\be
&&\Gamma_+(\lambda) = \Gamma_-(\lambda) (\1-2\pi i G_P(\lambda))\ ,\quad\quad \lambda\in\gamma:= \gamma_R\cup\gamma_L\cup i\bR\\
&&\Gamma(\lambda) = 
\le(\1 + \sum_{j=1}^{\infty} \frac {\Gamma_j}{\lambda^j}\ri),\quad\quad \lambda\longrightarrow\infty\\
&&G_P(\lambda):= 
				 \le[\begin{array}{ccccc}
					0 & (\epsilon_{2k_1}\;{\rm e}^{\Theta_{1}(\vec a_{1},\lambda)})^\T\chi_{\gamma} & (\epsilon_{2k_2}\;{\rm e}^{\Theta_{2}(\vec a_{2},\lambda)})^\T\chi_\gamma & \ldots & (\epsilon_{2k_n}\;{\rm e}^{\Theta_{n}(\vec a_{n},\lambda)})^\T\chi_\gamma\\
					&&&&\\
					{\rm e}^{-\Theta_{1}(\vec a_{1},\lambda)}\chi_{i\bR} & 0 & \ldots & \ldots & 0\\
					&&&&\\
					{\rm e}^{-\Theta_{2}(\vec a_{2},\lambda)}\chi_{i\bR} & \bD_{2,1}(\lambda) & \ddots & \vdots & \vdots\\
					&&&&\\
					\vdots & \vdots & \ddots & \ddots & \vdots\\
					&&&&\\
					{\rm e}^{-\Theta_{n}(\vec a_{n},\lambda)}\chi_{i\bR} & \bD_{n,1}(\lambda) & \ldots & \bD_{n,n-1} & 0
					\end{array}\ri]\nonumber\\
					&&\nonumber\\
		&&\bD_{i,j}(\lambda)=\chi_{i\bR}(\lambda)\left[(-1)^{t+1}{\rm e}^{(a_i^{(s)}-a_j^{(t)})\lambda+\frac{\tau_i-\tau_j}2\lambda^2}\right]_{s=1\ldots 2k_i, t=1\ldots 2k_j}\hspace{-15pt}\in\mathrm{Mat}(2k_i \times 2k_j)	
\ee
\end{problem}
\begin{remark}
In contrast to the Airy case here the jumps of the RH problem \ref{PRH} are already exponentially decaying and no deformations are needed.
\end{remark}
\begin{remark}
The same considerations as in Remark \ref{exAi} about the existence of the solutions for the RH problem \ref{PRH} apply here.
\end{remark}

The reader can verify that  $G_P(\lambda)=\f_P(\lambda)\g_P ^\T(\lambda)$ and that $\f_P^\T(\lambda) \g_P(\lambda)\equiv \0_n$ (see (\ref{defK_P3})) so that, similarly as in the previous section, we obtain the following theorem.
\begin{theorem}\label{FredholmtauP}
The Fredholm determinant $\det(\Id-\chi_{\vec I}P)$ equals to the  tau function $\tau_{JMU}$ in (\ref{taufunction}) related to the RH problem (\ref{PRH}). In particular $\forall\, i=1,\ldots, n$ and $\forall \,\ell=1,\ldots, 2k_i$ we have
\be
	&&\pa  \ln \det(\Id-\chi_{\vec I}P) = \omega_{(\1-2\pi i\, G_P)}(\partial):= \int_{\gamma}\mathrm{Tr}\Big(\Gamma_-^{-1}(\lambda )\Gamma'_-(\lambda)\Xi_{\partial }(\lambda )\Big)\frac{\d\lambda}{2\pi i}\nonumber\\
	&&\Xi_{\partial}  (\lambda ):= \le(\partial (\1-2\pi i\, G_P)\ri)(\1 - 2\pi i\, G_P)^{-1}= -2\pi i\, \pa G_P (\1 + 2\pi i\, G_P) \nonumber
\ee
where $\pa$ stands for any of the derivatives $\pa_{a_i^{(\ell)}},\ \pa_{\tau_\ell}$.
\end{theorem}
Again using the Miwa-Jimbo-Ueno residue formula we get some more explicit differential equations in terms of the coefficients of the symptotic expansion at $\infty$. As in the Airy case the jump matrix $G_P$ can be written as 
$
G_P(\lambda) = {\rm e}^{T_P(\lambda)} G_P^{(0)} {\rm e}^{-T_P(\lambda)}
$
where $G_P^{(0)}$ is a {\em constant} matrix (consisting of only $\pm 1$ and $0$) and 
\bea
&&T_P(\lambda) = {\rm diag}\le(T_P^{(0)},\vec T_P^{(1)},\dots, \vec T_P^{(n)}\ri),\qquad T_P^{(0)} := \frac{1}{1+\sum_{i=1}^n 2k_i} \sum_{i=1}^{n}\sum_{\ell=1}^{2k_i} \Theta_i(a_i^{(\ell)},\lambda)\label{TP}\\
&& \vec T_P^{(i)} := T_P^{(0)}\1-{\rm diag}\le(\Theta_i(a_i^{(1)},\lambda),\ldots,\Theta(a_i^{(2k_i)},\lambda)\ri)\ ,\qquad  \Tr \,\,T_P(\lambda)\equiv 0.
\eea
The matrix $\ds \Psi_P(\lambda):= \Gamma(\lambda)\, {\rm e}^{T_P(\lambda)} $
 solves a RHP with constant jumps and hence is (sectionally) a solution to a polynomial ODE.  As in the Airy case the integral in the Theorem \ref{FredholmtauP} can be converted in a formal residue, namely
\be
 \int_{\gamma}\mathrm{Tr}\Big(\Gamma_-^{-1}(\lambda)\Gamma'_-(\lambda)\Xi_{\partial}(\lambda)\Big)\frac{\d\lambda}{2\pi i} = -  \res_{\lambda=\infty} {\rm Tr} \le(\Gamma^{-1}(\lambda) \Gamma'(\lambda) \pa T_P \ri)\label{219bis}
\ee
We then find exactly as for Prop. \ref{Freddef}
\begin{prop}
\label{FreddefPearcey}
The Fredholm determinant $\det(\Id-\chi_{\vec I}P)$ satisfies
\bea
\pa_{a_i^{(\ell)}} \ln \det(\Id-\chi_{\vec I}P) =-(\Gamma_1)_{1+(\sum_{j<i}2k_j)+\ell,1+(\sum_{j<i}2k_j)+\ell}\\
\pa_{\tau_i} \ln \det(\Id-\chi_{\vec I}P) =\frac{1}2\sum_{\ell=1}^{2k_i}(\Gamma_1^2-2\Gamma_2)_{1+(\sum_{j<i}2k_j)+\ell,1+(\sum_{j<i}2k_j)+\ell}
\eea
where $\Gamma_1,\Gamma_2$ and $\Gamma(\lambda)$ are given by the solution of the RHP \ref{PRH}.
\end{prop}

\subsection*{Acknowledgements} 

The authors are grateful to A. Its who suggested  to try and apply the techniques used in \cite{BertolaCafasso1} to the present problem.

\appendix
\renewcommand{\theequation}{\Alph{section}.\arabic{equation}}

\section{Integrable kernels and isomonodromic tau functions}

In this appendix we review some basic facts about integrable kernels \cite{ ItsIzerginKorepinSlavnov} and, for sake of completeness, a theorem we originally proved in \cite{BertolaCafasso1} (see also \cite{BertolaCafasso2}).  
Given a piecewise smooth oriented curve ${\cal C}$ on the complex plane (possibly extending to infinity) and two matrix--valued functions  $\f,\g:\mathcal C\longrightarrow {\mathrm{Mat}_{p\times k}(\C)}$
we define the kernel $K$ as
$$K(\lambda,\mu):=\frac{\f^{\mathrm T}(\lambda)\g(\mu)}{\lambda-\mu}.$$ 
We say that such kernel is integrable if $\f^{\mathrm T}(\lambda)\g(\lambda) \equiv \0_k$ (so that it is non-singular). We are interested in the matrix--valued operator $K:L^2({\cal C},\C^k)\rightarrow L^2({\cal C},\C^k)$ acting on $k$-vector functions via the formula 
$$(Kh)(\lambda)=\int_{\cal C}K(\lambda,\mu)h(\mu)d\mu$$
and, in particular, we are interested in the Fredholm determinant $\det(\1-K)$ defined as in (\ref{FredhExp}). 
Denoting with $\partial$ the differentiation with respect to any auxiliary parameter on which $K$ may depend, we have Jacobi's formula
\be
	\partial \log\det(\1-K)=-\mathrm{Tr}((\Id+R)\partial K)
\label{resolvent}\ee 
where $R$ is the resolvent operator, defined as $R=(\1-K)^{-1}K$. Moreover $R$ is again an integrable operator
$$R(\lambda,\mu)=\frac{\mathbf F^{\mathrm T}(\lambda)\mathbf G(\mu)}{\lambda-\mu}$$ where  $\mathbf F,\mathbf G$ are given by 
$  \mathbf F(\lambda):=\Gamma(\lambda)\mathbf f(\lambda)$, $ 
		\mathbf G(\lambda) :=(\Gamma^{-1})^{\mathrm T}(\lambda)\mathbf g(\lambda)$, and $\Gamma$ is the solution of the following RH problem:
\bea 
		\Gamma_+(\lambda)&\&=\Gamma_-(\lambda)M(\lambda)\quad\lambda\in{\cal C}\label{RHA1}\\
		\Gamma(\lambda)&\& \sim\1+\mathcal O(\lambda^{-1})\quad\lambda\longrightarrow\infty\label{RHA2}\\
	M(\lambda)&\& =\1-2\pi i\f(\lambda)\g^{\mathrm T}(\lambda)\label{RHA3}
	\label{RH}
\eea
The  theory also guarantees that the solution to the above problem exists if and only if the Fredholm determinant does not vanish.
Now suppose that the kernel the matrices $\f, \g$ and thus $K$ (and hence the Riemann--Hilbert problem (\ref{RHA1})--(\ref{RHA3})) depend smoothly on  parameters. On the space of these deformation parameters, we introduce the following one-form\footnote{Here and below we will denote with $'$ the derivative with respect to $\lambda$.}
\bea
	\omega_M(\partial)&\&:= \int_{\cal C}\mathrm{Tr}\Big(\Gamma_-^{-1}(\lambda)\Gamma'_-(\lambda)\Xi_\partial(\lambda )\Big)\frac{\d \lambda}{2\pi i}
	\label{omegamalgrange}\ ,\qquad
	\Xi_\partial(\lambda):=\partial M(\lambda)M^{-1}(\lambda).
\eea
Moreover, in the case that $\omega_M$ (which is a condition only on the jump matrices as seen in \cite{BertolaIsoTau} is closed in the space of the deformation parameters, it is also defined (at least locally) the isomonodromic tau function $\tau_{JMU}$ such that
\be
	\partial\log\tau_{JMU} := \omega_M(\partial)\label{taufunction}
\ee 
for every vector field $\partial$ in the space of deformation parameters.
The definitions (\ref{omegamalgrange}),(\ref{taufunction}) are posed for arbitrary jump matrices; in the case of the RHP (\ref{RHA1})--(\ref{RHA3}) $\tau_{JMU}$ in (\ref{taufunction}) and the Fredholm determinant are related as in the theorem below.
\begin{theorem}[\cite{BertolaCafasso1}]\label{thdetMalgrange}\footnote{Actually the article \cite{BertolaCafasso1} treats the case $k=1$, but the proof does not change considering this more general case.}
Let $\f(\lambda;\vec s), \g(\lambda;\vec s): \C \times S \longrightarrow {\mathrm{Mat}_{p\times k}(\bC)}$ and consider the RHP with jumps as in (\ref{RHA1})--(\ref{RHA3}).
Given any vector field $\partial$ in the space of the parameters $S$ of the integrable kernel we have the equality
	\be
		\omega_M(\partial)=\partial\ln\det(\1-K) + \int_{\cal C} \pa \,\mathrm{Tr}\Big({ \f'}^\T \g\Big)d\lambda  + 2\pi i\int_{\cal C}\mathrm{Tr}( \g^\T \f' 
\partial\g^\T\f) d\lambda
	\label{detMalgrange}\ee
	where $\omega_M(\partial)$ is as in (\ref{omegamalgrange}). 
\end{theorem}
In the cases we treat in this article the additional terms in (\ref{detMalgrange}) vanish because of the particular shape of the matrices $\f, \g$: indeed, due to their particular shape, a stronger condition holds that $\f^\T(\lambda)\g(\mu) \equiv \0_k$  when $\lambda$ and $\mu$ both belong to the same contour.
Hence the tau function $\tau_{JMU}$, defined in (\ref{taufunction}), coincides with the Fredholm determinant $\det(\1-K)$.

\section{A brief reminder of regularized Fredholm determinants}
\label{appregdet}
We refer to \cite{Si} for the relevant details: we shall need only the elementary facts which we recall here. In general the Fredholm determinant of an operator of the form $(\Id - G)$ can be defined only when $G$ is of trace class. If $G$ is represented as an integral operator on a (separable) Hilbert  $L^2(X,\d\mu)$ with kernel $G(\xi,\eta)$ (we abuse notation by using the same symbol for the operator and  kernel) then\footnote{Note that, if $X=\lbrace 1,\ldots, n\rbrace\times S$ and $\mu=\nu\otimes \lambda$ where $\nu$ is the counting measure, the formula above covers the case of matrix--valued kernels on $S$ equipped with measure $\lambda$, as in \cite{JoDetProc}.}
\be
\det(\Id  - G) := 1 + \sum_{k=1}^\infty \frac 1 {k!} \int_{X^k} \det [G(\xi_i,\xi_j)]_{i,j\leq k}\prod_{i=1}^k \d\mu(\xi_i)
\label{FredhExp}\ee
The trace ideals $\mathcal I_p$, $p\in \mathbb N$ consist of operators whose $p$-th power is trace-class \cite{Si}; in particular $\mathcal I_2$ consists of Hilbert-Schmidt operators.   For $G\in \mathcal I_p$ one can define following Carleman a regularized determinant $\det_p(\Id -G)$ which has the same main property of vanishing iff the operator is not invertible. In particular for Hilbert--Schmidt operators one has 
\be
\mathrm{det}_2(\Id - G):= 1 + \sum_{k=1}^\infty \frac 1 {k!} \int_{X^k} \det [G(\xi_i,\xi_j) (1-\delta_{ij})]_{i,j\leq k}\prod_{i=1}^k \d\mu(\xi_i)
\ee
that is, we  omit the diagonal elements in the determinants under the integral sign. It satisfies
\begin{itemize}
\item if $G \in \mathcal I_1 \cap \mathcal I_2$  then 
$\ds 
\mathrm{det}_2(\Id - G) = \det (\Id - G) {\rm e}^{{\Tr} G}
$
\item 
if $G_1,G_2$ are Hilbert--Schmidt operators (and hence $G_1G_2$ is trace class) then
\be
\mathrm{det}_2(\Id - G_1) \mathrm{det}_2 (\Id-G_2) = \mathrm{det}_2(\Id - G_1 - G_2 + G_1G_2) {\rm e}^{{\Tr} (G_1G_2)}
\label{det_2formula}\ee 
\end{itemize}
An interesting occurrence (which is used in this article) is that if $G$ is just HS but its kernel vanishes on the diagonal $G(\xi,\xi)\equiv 0$ then the series defining $\det_2(\Id-G)$ is identical to the regular $\det(\Id- G)$. The reason for still wanting to distinguish $\det_2$ from $\det$ in this case is simply that $G$ may fail to have a trace and in a different basis of the Hilbert space the ordinary $\det$ may simply be ill--defined.

\bibliographystyle{plain}
\bibliography{/Users/bertola/Documents/Papers/BibDeskLibrary.bib}
 \end{document}